\documentclass [preprint,prb,aps,letterpaper,showpacs,showkeys,onecolumn,amsmath,amssymb,floatfix] {revtex4}
\usepackage{graphicx}
\usepackage{color}
\usepackage{textcomp}
\usepackage{mathptmx} 

\begin{document}
\vskip 0.4cm

\title{ Electronic transport in ferromagnetic barriers on the surface of a topological insulator with $\delta$ doping}

\author{ Jian-Hui Yuan$^{a}$}
\altaffiliation{The corresponding author, E-mail:
jianhui831110@163.com. Tel : 011+150-7882-3937}
\author{ Yan Zhang$^{a}$}
\author{ Daizheng Huang$^{a}$}
\author{ Qinhu Zhong$^{b}$}
\author{ Xin Zhang$^{c}$}
\affiliation{$^{a}$The department of Physics, Guangxi medical
university, Nanning, Guangxi, 530021, China \\ $^{b}$School of
Physics, School of  Physics and Optical information sciences, Jia
Ying university, Meizhou, Guangdong 514015, China\\$^{c}$The 34th
Institute of China Electronics and Technology Group Corporation,
Guilin Guangxi 541004, China}
\date{\today}



\begin{abstract}
{\qquad\small {\bf Abstract:}  {We investigate electron transporting
through a two-dimensional ferromagnetic/normal/ferromagnetic tunnel
junction on the surface of a three-dimensional topological insulator
with taking into $\delta$ doping account. It is found that the
conductance oscillates with the Fermi energy, the position and the
aptitude of the $\delta$ doping. Also the conductance depends
sensitively on the direction of the magnetization of the two
ferromagnets, which originate from the control of  the spin flow due
to spin-momentum locked. It is found that the conductance is the
maximum at the parallel configuration while it is minimum at the
antiparallel configuration and vice versa, which  may stem from the
half wave loss due to the electron wave entering through the
antiparallel configuration. These characters are very helpful for
making new types of magnetoresistance devices due to the practical
applications.
 }}
\end{abstract}

\maketitle
\section{\textbf{Introduction}}

The concept of a topological insulator (\textbf{TI}) dates back to
the work of Kane and Mele, who focused on the two-dimensional
(\textbf{2D}) systems
[1]. 
 Its  discovery in
theoretical [2] and experimental [3] has accordingly generated a
great deal of excitement in the condensed matter physics community.
 Recent theoretical and experimental discovery of the \textbf{2D} quantum spin Hall system [4-13] and its generalization
to the topological insulator in three dimensions [14-16] have
established the state of matter in the time-reversal symmetric
 systems. Surface sensitive experiments such as angle-resolved
photoemission spectroscopy (\textbf{ARPES}) and scanning tunneling
microscopy (\textbf{STM}) [1,2] have confirmed the existence of
this exotic surface metal, in its simplest form, which takes a single Dirac dispersion. 

 Topological insulator  is a new state of matter,
distinguished from a regular band insulator by a nontrivial
topological invariant, which characterizes its band structure
protected by time-reversal invariant[1,2,13,17,18].  There has been
much recent interest in topological insulators (\textbf{TIs}),
three-dimensional insulators with metallic surface states. In
particular, the surface of a three-dimensional (\textbf{3D}) TI,
such as Bi$_{2}$Se$_{3}$ or Bi$_{2}$Te$_{3}$ [17], is a 2D metal,
whose band structure consists of an odd number of Dirac cones,
centered at time reversal invariant momenta in the surface Brillouin
zone [18].  This corresponds to the infinite mass Rashba model [19],
where only one of the spin-split bands exists. This has been
beautifully demonstrated by the spin- and angle-resolved
photoemission spectroscopy [20,21].  On the one hand,  the
\textbf{3D} \textbf{TIs} are expected to show several unique
properties when the time reversal symmetry is broken [22-24]. This
can be realized directly by a ferromagnetic insulating (\textbf{FI})
layer attached to the 3D TI surface  with taking into the proximity
effect account. On the other hand, The topological surface states
may be applied to the spin field-effect transistors in spintronics
due to strong spin-orbit coupling [25,26], such as giant
magnetoresistance[27] and tunneling magnetoresistance[28,29,30] in
the metallic spin valves. Wang et al[27] investigated room
temperature giant and linear magnetoresistance in topological
insulator, which is useful for practical applications in
magnetoelectronic sensors such as disk reading heads, mechatronics,
and other multifunctional electromagnetic applications. In Ref.28
Xia et al studied anisotropic magnetoresistance in topological
insulator, which can be explained as a giant magnetoresistance
effect. Kong et al[29] predicted a giant magnetoresistance as large
as 800\% at room temperature with the proximate exchange energy of
40 meV at the barrier interface. Yokoyama et al[30] investigated
charge transport in two-dimensional ferromagnet/ferromagnet junction
on a topological insulator. Their results are given in the limit of
thin barrier, which can be view as $\delta$ potential barrier due to
the mismatch effect and built-in electric field of junction
interface. In the meantime, the transport property of the
topological metal (\textbf{TM}) have been attracted a lot of
attention[31-34]. References 31, 32, 33 and 34 investigated electron
transport through a ferromagnetic barrier on the surface of a
topological insulator, such as electron tunneling, tunneling
magnetoresistance and spin valve. In these papers, a remarkable
feature of the Dirac fermions is that the Zeeman field acts like a
vector potential, which is in contrast to the Schr$\ddot{o}$dinger
electrons in conventional semiconductor heterostructures modulated
by nanomagnets. However, there is a few papers to investigate the
$\delta$ doping effect on the electronic transport on the surface of
a TI.

In this work, we study the transport properties in a \textbf{2D}
ferromagnet/normal/ferromagnet junction on the surface of a strong
topological insulator where a $\delta$ doping potential is exerted
on the normal segment. As shown in Fig.1, for the ferromagnetic
barrier, a ferromagnetic insulator is put on the top of the
\textbf{TI} to induce an exchange field via the magnetic proximity
effect. So far such a system has not been well studied.  We find
that the conductance oscillates with the Fermi energy, the position
and the aptitude of the $\delta$ doping. Also the conductance
depends sensitively on the direction of the magnetization of the two
ferromagnets, which originate from the control of  the spin flow due
to spin-momentum locked.  It is found that the conductance is the
maximum at the parallel configuration while it is minimum at the
antiparallel configuration and vice versa, which shows a control of
giant magnetoresistance (\textbf{MR}) effect. In order to interprets
this phenomena, we propose  a hypothesis that the half wave loss in
the case of the electron wave entering through the antiparallel
configuration.  These characters are very helpful for making new
types of magnetoresistance devices due to the practical
applications. This paper is organized as follows. In Sec.
\uppercase\expandafter{\romannumeral 2} , we introduce the model and
method for our calculation. In Sec.
\uppercase\expandafter{\romannumeral 3}, the numerical analysis to
our important analytical issues are reported. Finally, a brief
summary is given in sec. \uppercase\expandafter{\romannumeral 4}.

\section{ model and method}
\begin{figure}
\includegraphics[scale=0.8]{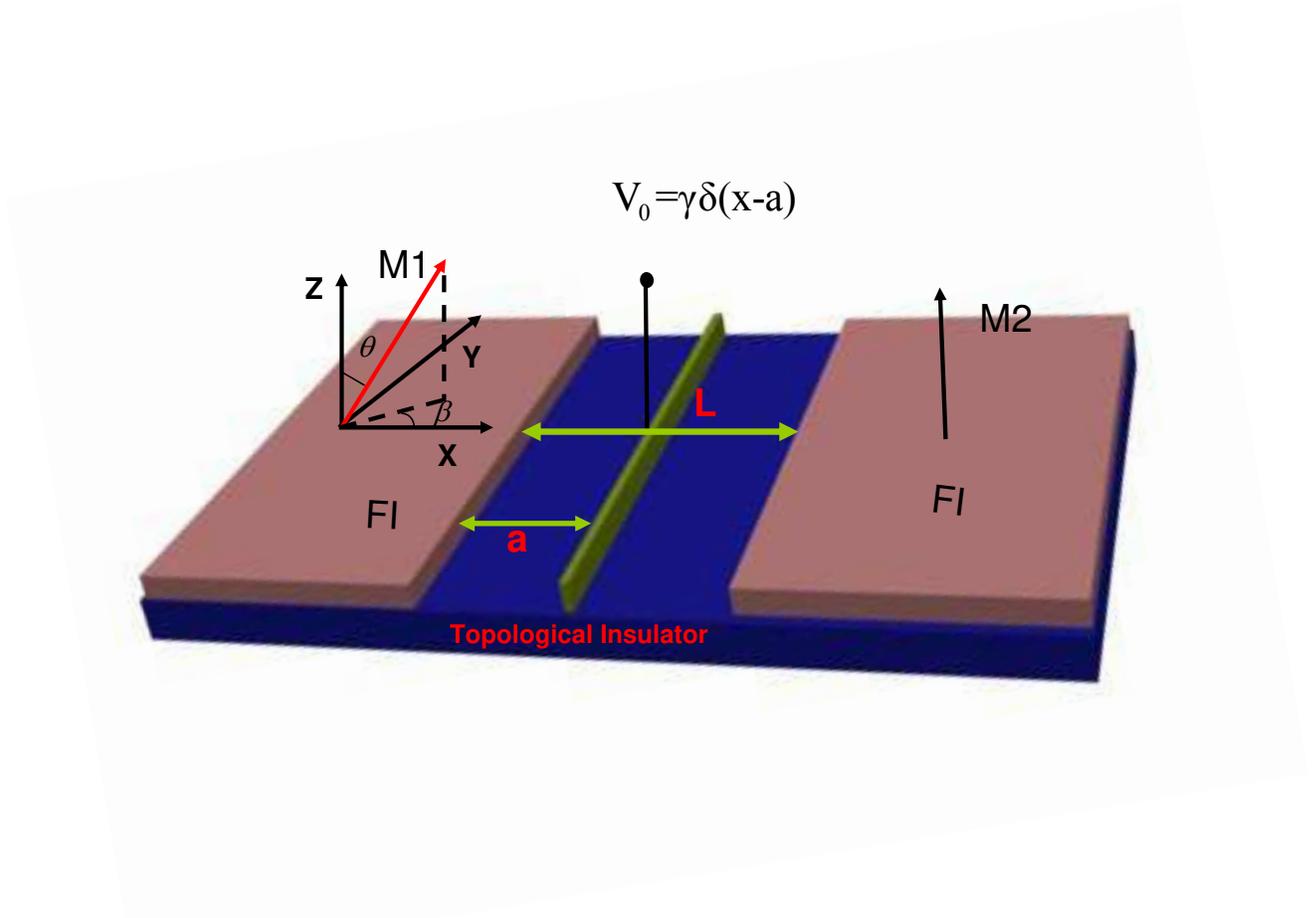}
\caption{  Schematic layout of a 2D ferromagnet/ normal/ferromagnet
junction on the surface of a topological insulator with taking into
$\delta$ account.The ferromagnetism is induced in the topological
surface state due to the proximity effect by the ferromagnetic
insulators deposited on the top, and the central normal segment is
insert a $\delta$ potential } \label{FIG.1.eps}
\end{figure}
Now, let us consider a ferromagnetic/normal/ferromagnetic tunnel
junction which is deposited on the top of a topological surface
where a $\delta$ doping potential is exerted on the normal segment.
The ferromagnetism is induced due to the proximity effect by the
ferromagnetic insulators deposited on the top as shown in Fig. 1.
The bulk \textbf{FI} interacts with the electron in the surface of
\textbf{TI} by the proximity effect, which is induced a
ferromagnetism in the surface of \textbf{TI}. Thus we focus on
charge transport at the Fermi level of the surface of \textbf{TIs},
which is described by the \textbf{2D} Dirac Hamiltonian
\begin{eqnarray}
H = \upsilon_{F} \sigma \cdot \textbf{P} + \sigma\cdot M + V{(x)},
\end{eqnarray}\label{1}
where  $\sigma$ is Pauli matrices , the $\delta$ potential is
exerted on the normal segment with $V(x)=\gamma \delta(x-a)$ where
$a$ is the position of the $\delta$ potential. In our model, we
choose the effective exchange field in the left region with
\textbf{$M_1$}=$
(m_x,m_y,m_z)=m_1(\sin\theta\cos\beta,\sin\theta\sin\beta,\cos\theta)$.
We assume that the initial magnetization of \textbf{ FI} stripes in
the right region  is aligned with the +z axis,
\textbf{$M_2$}=$(0,0,m_2)$.  In an actual experiment, one can use a
magnet with very strong (soft) easy axis anisotropy to control the
ferromagnetic. Because of the translational invariance of the system
along $y$ direction, the equation $H \Psi(x, y)=E \Psi(x, y)$ admits
solutions of the form $\Psi(x,y)=(\Psi_{1}(x),
\Psi_{2}(x))^{T}\exp(i k_y y)$. We set $\hbar = \upsilon_F =L=1$ in
the following where $L$ is the distance between the two bulk
\textbf{FI} as a unit of the length, so the unit of the energy is
given with the form $E_0=\hbar\upsilon_F/L$. To simplify the
notation, we introduce the dimensionless units: $\vec{r}\rightarrow
\vec{r} L$ , $E \rightarrow E E_0$. Due to the presence of the
ferromagnet, time-reversal symmetry is broken which can lead to the
robustness against disorder. Here, we assume that $L$ is shorter
than the mean-free path as well as the spin coherence length, so we
can ignore the disorder. In order to investigate the $\delta$
doping, we describe the length $b \sim 0$ of the barrier with the
potential $U\sim\infty$ while keeping $\gamma\equiv U b$ =const to
replace the $\delta$ potential.

With the above Hamiltonian, wave function in whole system is given
by
\begin{eqnarray}
\Psi_{1}&=&\left\{\begin{array}{lllll  }  A \exp [i (k_{1}-m_x) x] + \\
B \exp [-i (k_1+m_x) x], &
\quad x<0, \\
C_1 \exp (i k x) + D_1 \exp (i k x), & \quad 0<x<a,
\\
C_2 \exp (i q x) + D_2 \exp (i q x), & \quad a<x<a+b,
\\
C_3 \exp (i k x) + D_3 \exp (i k x), & \quad a+b<x<1,
\\
F\exp (i k_2 (x-1)) , & \quad x>1,\end{array} \right.
\end{eqnarray}\label{2}

\begin{eqnarray}
\Psi_{2}&=&\left\{\begin{array}{lllll  }  A \delta_1 \frac{k_1+ i (k_y+m_y)}{\triangle_1}\exp [i (k_{1}-m_x) x] +\\
 B \delta_1 \frac{-k_1+ i (k_y+m_y)}{\triangle_1}\exp [-i (k_1+m_x) x],
&
\quad x<0, \\
C_1 \frac{k+i k_y}{E} \exp (i k x) + D_1 \frac{-k+i k_y}{E} \exp (i
k x), & \quad 0<x<a,
\\
C_2 \frac{q+i k_y}{E-U} \exp (i q x) + D_2 \frac{-q+i k_y}{E-U}\exp
(i q x), & \quad a<x<a+b,
\\
C_3 \frac{k+i k_y}{E} \exp (i k x) + D_3 \frac{-k+i k_y}{E} \exp (i
k x), & \quad a+b<x<1,
\\
F\delta_2 \frac{k_2+ i ky}{\triangle_2}\exp (i k_2 (x-1)) , & \quad
x>1,\end{array} \right.
\end{eqnarray}\label{2}
where $\triangle_1=\sqrt{E^2-m_z^2}=\sqrt{k_1^2+(k_y+m_y)^2}$,
$\delta_1=(E-m_z)/\triangle_1$,
$\triangle_2=\sqrt{E^2-m_2^2}=\sqrt{k_2^2+k_y^2}$,
$\delta_2=(E-m_2)/\triangle_2$, and $k_{1}=\triangle_1\cos
{\phi_{F_1}}$, $k_{2}=\triangle_2\cos {\phi_{F_2}}$, $
k_y=\triangle_1\sin {\phi_{F_1}}-m_y=\triangle_2\sin {\phi_{F_2}}$
wave vectors in left region  and in the right region, respectively.
$k=\sqrt{E^2-k_y^2}$, $q=\sqrt{(E-U)^2-k_y^2}$ in the the normal
segment. The momentum $k_y$ conservation should be satisfied
everywhere. Continuities of the wave function  $\Psi$ at $x=0, a,
a+b$ and $x=1$, wave functions are connected by the boundary
conditions:
\begin{eqnarray}
\Psi(0^{-}) = \Psi(0^{+}), \Psi(a^{-}) = \Psi(a^{+}), \nonumber\\
\Psi((a+b)^{-}) = \Psi((a+b)^{+}), \Psi(1^{-}) = \Psi(1^{+}),
\end{eqnarray}\label{5}
which determine the coefficients A,B,$C_1,D_1,C_2,D_2,C_3,D_3$ and F
in the wave functions.

In the liner transport regime and for low temperature, we can obtain
the conductance $G$ by introducing it as the electron flow averaged
over half the Fermi surface from the well-known Landauer-Buttiker
formula$^{25, 29, 30,34}$, it is straightforward to obtain the
ballistic conductance G at zero temperature
\begin{eqnarray}
G=\frac{e^2 W_y E_F}{\pi h}\frac{1}{2}\int_{-\pi/2}^{\pi/2}
d\phi_{F_1}\frac{F^* F}{A^*
A}\frac{\delta_2\Delta_1}{E_F\delta_1}\cos\phi_{{F_2}}
\end{eqnarray}\label{5}
where $W_y$ is the width of interface along the y direction, which
is much larger than L($L\equiv1$), and we take E as $E_F$ , because
in our case the electron transport happens around the Fermi level.

\section{Results and Discussions}
\begin{figure}
\includegraphics[scale=0.8]{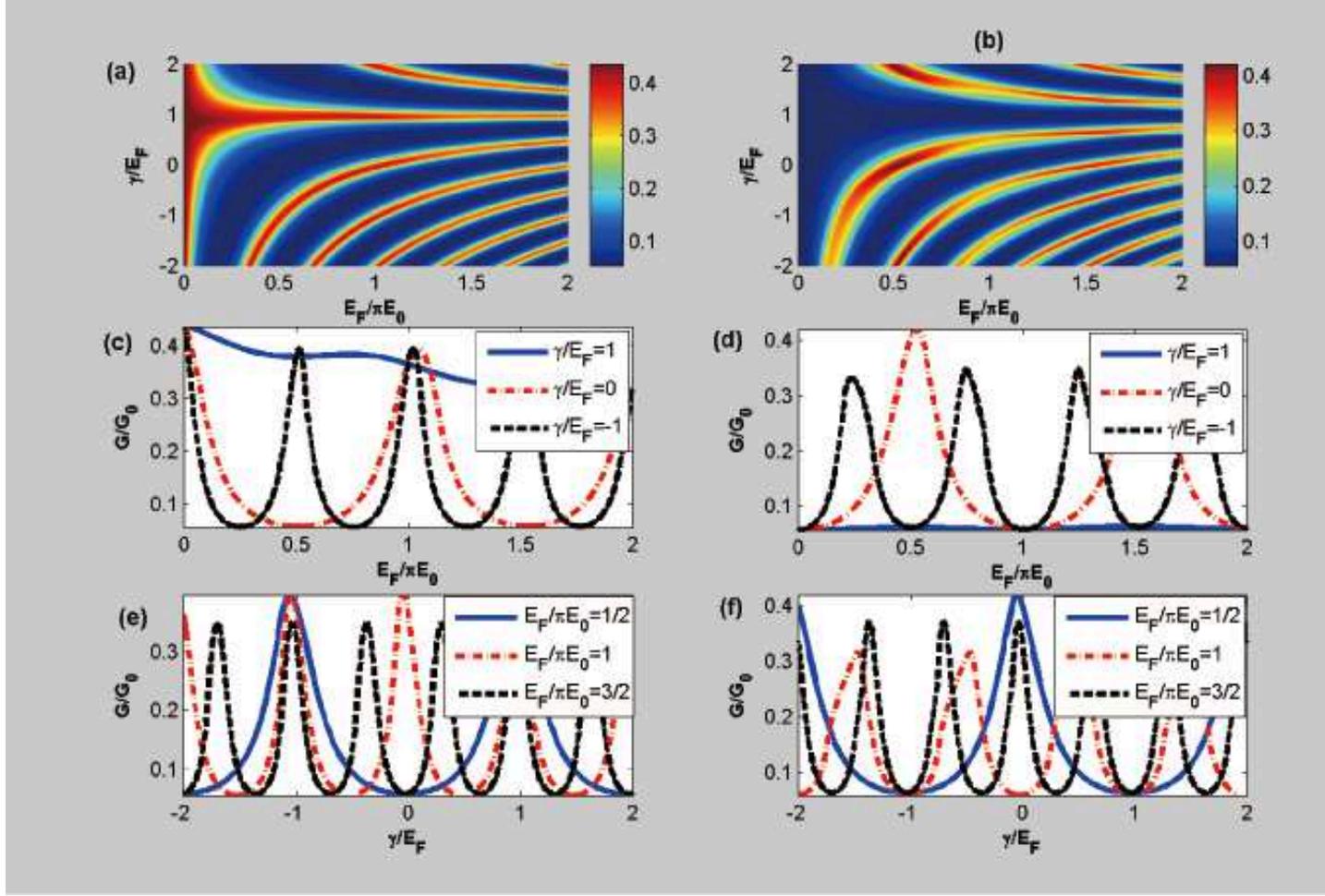}
\caption{  Schematic diagram of the normalized conductance $G/G_0$
as a function of the Fermi energy $E_F/\pi E_0$ and the doping
potential parameter $\gamma/E_F$ for $m_z=m_2=0.9E_F$ and $a=0.5$.
The left panels (a), (c) and (e) correspond to the parallel
configuration and the right panels (b), (d) and (f) correspond to
the antiparallel configuration.} \label{FIG.2.eps}
\end{figure}
In what follows, we use $G_0\equiv \frac{e^2 W_y E_F}{\pi h}$ as the
unit of the conductance. It is worth noting that we set $\hbar =
\upsilon_F =L=1$ where $L$ is the distance between the two bulk
\textbf{FI} as a unit of the length, so the unit of the energy is
given with the form $E_0=\hbar\upsilon_F/L$.

In Fig.2, we show the normalized conductance $G/G_0$ as a function
of the Fermi energy $E_F/\pi E_0$ and the doping potential parameter
$\gamma/E_F$ for $m_z=m_2=0.9E_F$ and $a=0.5$. The left panels (a),
(c) and (e) correspond to the parallel configuration and the right
panels (b), (d) and (f) correspond to the antiparallel
configuration. It is easily seen that the conductance oscillates
with Fermi energy $E_F/\pi E_0$ and the doping potential parameter
$\gamma/E_F$. The difference of the conductance  is very obvious
between the negative doping potential parameter $\gamma/E_F$ and the
positive doping potential parameter $\gamma/E_F$.  It is easily seen
that the oscillation period of the conductance against the Fermi
energy $E_F/\pi E_0$ corresponding to a negative doping potential
parameter $\gamma/E_F$ is smaller than that corresponding to the
same value but positive doping potential parameter $\gamma/E_F$. It
is found that the conductance is the maximum at the parallel
configuration while it is minimum at the antiparallel configuration
and vice versa, which shows a control of giant magnetoresistance
effect [see Fig 2(a) and (b)] as the similar  report by Ref.
[27,30]. In order to further investigate these effect, we fix the
barrier potential parameter $\gamma/E_F$ to discuss the electronic
conductance against the Fermi energy $E_F/\pi E_0$ as shown in Fig
2(c) and (d), where the panels (c) and (d) correspond to the
parallel configuration and the antiparallel configuration,
respectively. We can see that the conductance oscillates with
increasing the Fermi energy $E_F/\pi E_0$ due to the phase factor $k
a$ and $k L$ ($L\equiv1$) from Eq. (3), where $k$ is the function as
the the Fermi energy $E_F/\pi E_0$.  These oscillate states may
originate from the electron confined in the region between the
ferromagnetic segment and $\delta$ doping segment. Also we can
easily see that $\delta$ doping change dramatically the conductance.
The conductance changes with the Fermi energy $E_F/\pi E_0$ in the
same way  both in Fig. 2(c) and (d). The difference is that the
conductance is maximum in Fig. 2(c) while it is minimum in Fig.
2(d), and vice versa. It is more interesting to us, the conductance
as the function of the Fermi energy $E_F/\pi E_0$ is compressed
totally in the antiparallel configuration compared with the parallel
configuration, which shows a quantum switch on-off property.  In
Fig.2 (e) and (f), we discuss the change of the electronic
conductance against the doping potential parameter $\gamma/E_F$,
where the panels (e) and (f) correspond to the parallel
configuration and the antiparallel configuration, respectively.  The
transmission coefficient is rather complicated but we note that it
contains the doping parameter $\gamma$ only in the form of $\cos
\gamma$ and $\sin \gamma$, and the transmission probability in a
certain
 Fermi energy $E_F/\pi E_0$ can write a general formula as $T(\gamma)=\alpha_0+(\cos 2\gamma-\alpha_1) /[\alpha_2 \cos 2\gamma + \alpha_3 \sin 2\gamma+\alpha_4]$,
 where all of $\alpha_i (i=0,1,2,3,4)$ is the function of  the Fermi energy $E_F/\pi E_0$ and $k_y$. We note the fact that $T(\gamma)\neq T(-\gamma)$, thus we easily
 understand the fact that the conductance for $\gamma/E_F=1$ is not the same to that for $\gamma/E_F=-1$ unless the Fermi energy $E_F/ E_0\equiv m \pi$ where $m$ is a integer [see in Fig.2 (c) and (d)]. However, we can note that the
 transmission probability and hence the conductance are $\pi$ periodic with respect to $\gamma$. It is easily seen that the change of conductance between maximum
 and minimum by the $\delta$ doping is similar to the spin field-effect transistor, where the modulation of the conductance arises from the spin precession due to
 spin-orbit coupling [25]. The reason is that the spin direction of wave function rotates through the different regions due to spin-momentum locked. Furthermore,
 we also note that the conductance is maximum in Fig. 2(e) while it is
minimum in Fig. 2(f), and vice versa. It is give us a chance to
obtain a large maximum/minimum ratio of the conductance, which is
important for a transistor.

\begin{figure}
\includegraphics[scale=0.8]{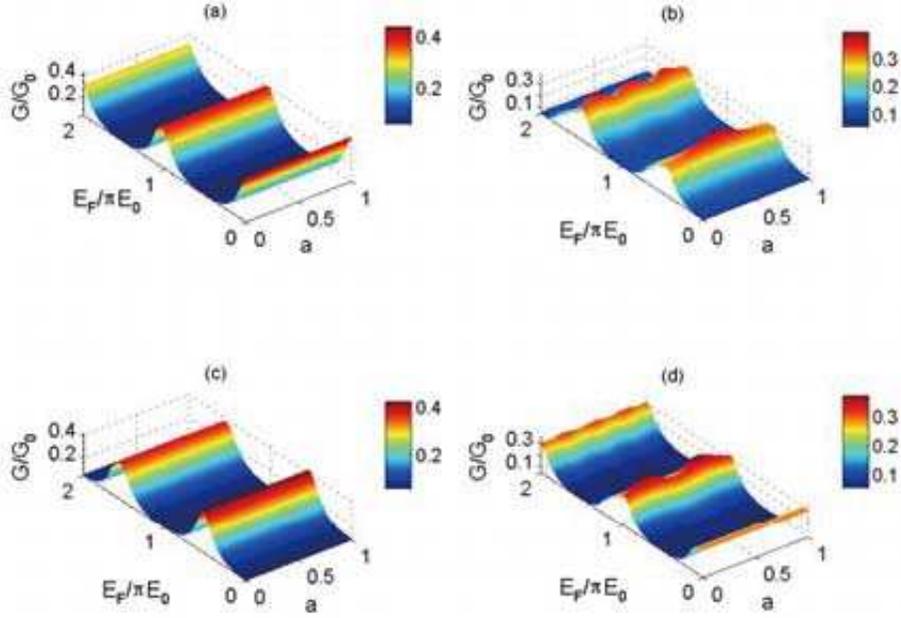}
\caption{ Schematic diagram of the normalized conductance $G/G_0$ as
a function of the Fermi energy $E_F/\pi E_0$ and the doping position
parameter $a$ for $m_z=m_2=0.9E_F$  for the  different doping
potential parameters [(a) and (c)] $\gamma/\pi=0$ and [(b) and (d)]
$\gamma/\pi=1/2$. The panels (a) and (b) correspond to the parallel
configuration and the panels (c) and (d) correspond to the
antiparallel configuration. }\label{FIG.3.eps}
\end{figure}

\begin{figure}
\includegraphics[scale=0.8]{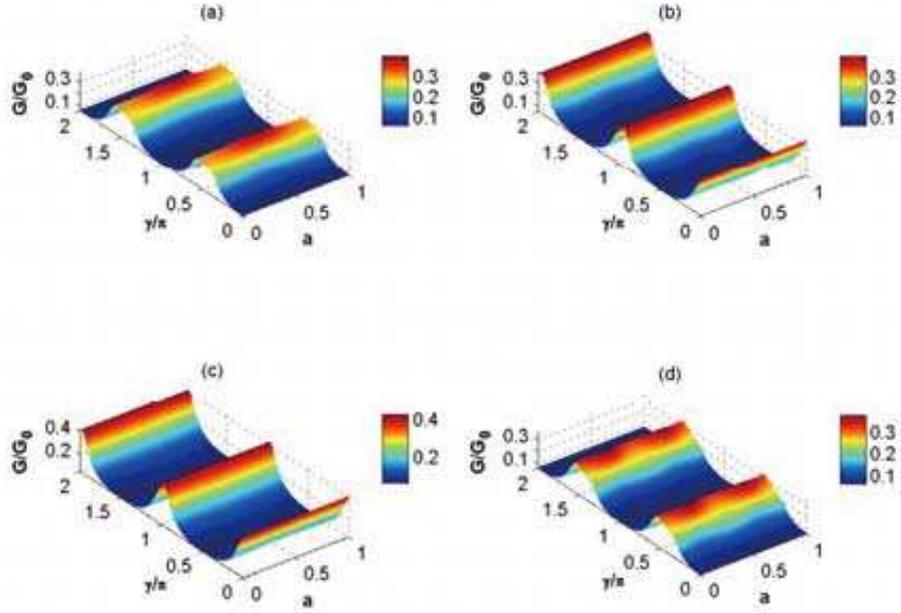}
\caption{Schematic diagram of the normalized conductance $G/G_0$ as
a function of the the doping potential $\gamma$ and the doping
position parameter $a$ for $m_z=m_2=0.9E_F$ for the  different Fermi
energy [(a) and (c)] $E_F/\pi E_0=1/2$ and [(b) and (d)] $E_F/\pi
E_0=1$ . The panels (a) and (b) correspond to the parallel
configuration and the panels (c) and (d) correspond to the
antiparallel configuration. }\label{FIG.4.eps}
\end{figure}

In order to investigate the effect of the position of the $\delta$
doping on the conductance, we show the normalized conductance
$G/G_0$ in Fig.3 as a function of the Fermi energy  $E_F/\pi E_0$
and the doping position parameter $a$ for $m_z=m_2=0.9E_F$ for the
different doping potential parameters [(a) and (c)] $\gamma/\pi=0$
and [(b) and (d)] $\gamma/\pi=1/2$. The panels (a) and (b)
correspond to the parallel configuration and the panels (c) and (d)
correspond to the antiparallel configuration. It is easily seen that
for the fixed $\delta$ doping potential parameters $\gamma$ the
conductance is maximum in parallel configuration [ see in Fig. 3(a)
and (b)] while it is minimum in antiparallel configuration [see in
Fig. 3(c) and (d)], and vice versa. For $\delta$ doping potential
parameters $\gamma/\pi=0$,
 there is no doping in the normal segment. Thus  the conductance is not change by changing the doping position parameter $a$. It is more interesting
 to us that the conductance is $\pi$ periodic with respect to the Fermi energy $E_F/\pi E_0$. The reason is that there are some electron state in the
 normal segment where the electron is confined into it. When the incident wave length in the normal segment satisfies $L\equiv1=m \lambda$ where $m$ is
 integer, quantum interference in the normal segment can enhance the transmitted wave.  That is to say, there are some confined state, $E\geq k = m \pi$.
 When this condition above is satisfied, a peak of conductance can appear. For the antiparallel configuration, we can assume that there is a half
 wave loss when the incident wave length in the normal segment reflect many times between the two different ferromagnetic layer, which is analogy to the
 optics reflected in different medius.  That is to say, there are some confined state, $E\geq k = (m+1/2) \pi$. When this condition above is satisfied, a
 peak of conductance will appear. Thus there is $1/2 \pi$ energy difference between  the Fermi energy corresponding to the peak of the conductance in parallel
 configuration and that in antiparallel configuration. When the $\delta$ doping
 appears, the quantum interference condition is change, $E\geq k = m
 \pi-\gamma$. So we can easily understand that the conductance is
 maximum in Fig. 3(a) and (c) where conductance is
 minimum in Fig. 3(b) and (d) and vice versa. For antiparallel
 configuration, there is also a half
 wave loss. Furthermore, we can find that the position of $\delta$
 doping in the normal segment can affect the electronic conductance because of the position of $\delta$
 doping also induced quantum phase interference in a certain Fermi
 energy. It is worth noting that a large the Fermi energy mean to a large of the number of confined
 electron state. Thus, the number of peaks for the large Fermi energy is more than that for the small Fermi energy .
 The similar results are seen in Fig.4. It is reasonable to us that $\pi$ periodic of the conductance appears as the above
 analysis. Compared with the parallel configuration, there is also a
 half-wave loss when the electron wave enter through the antiparallel
 configuration.

\begin{figure}
\includegraphics[scale=0.8]{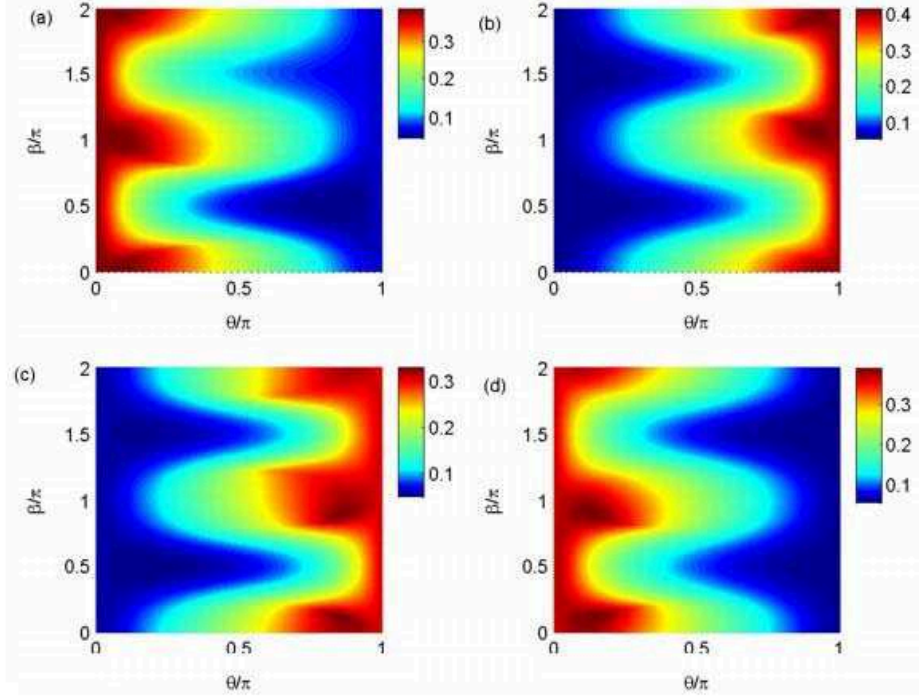}
\caption{Schematic diagram of the normalized conductance $G/G_0$
with $m_z=m_2=0.9E_F$ and $a=0.5$ for (a) $\gamma/\pi=1/2$ and
$E_F/\pi E_0=1/2$, (b) $\gamma/\pi=0$ and $E_F/\pi E_0=1/2$, (c)
$\gamma/\pi=1/2$ and $E_F/\pi E_0=1$, and (d) $\gamma/\pi=0$ and
$E_F/\pi E_0=3/2$. }\label{FIG.5.eps}
\end{figure}



\begin{figure}
\includegraphics[scale=0.8]{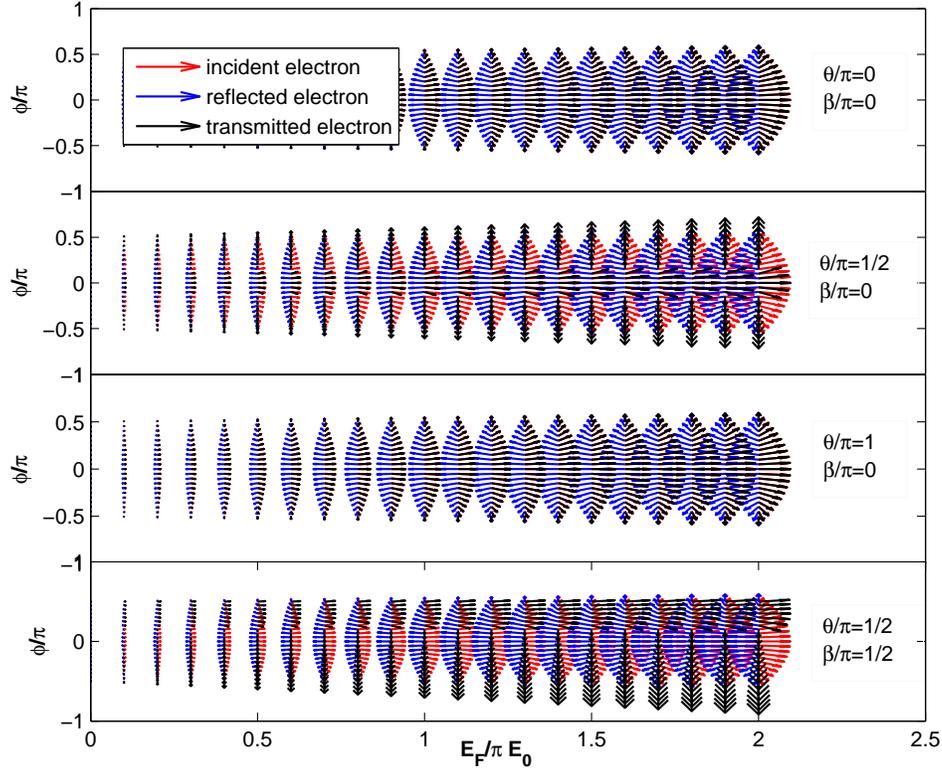}
\caption{Schematic diagram of the spin orientation of incident( red
arrows), transmitted (black arrows) and reflected (blue arrows)
electron as a function of the Fermi energy $E_F/\pi E_0$ and the
incident angle $\phi$ with $m_z=m_2=0.9E_F$ and  $a=0.5$ for the
four cases (a) $\theta/\pi=0, \beta/\pi=0$, (b) $\theta/\pi=1/2,
\beta/\pi=0$, (c) $\theta/\pi=1, \beta/\pi=0$. and
(d)$\theta/\pi=1/2, \beta/\pi=1/2$}\label{FIG.8.eps}
\end{figure}

In Fig.5, we show schematic diagram of the normalized conductance
$G/G_0$ with $m_z=m_2=0.9E_F$ and $a=0.5$ for (a) $\gamma/\pi=1/2$
and $E_F/\pi E_0=1/2$, (b) $\gamma/\pi=0$ and $E_F/\pi E_0=1/2$, (c)
$\gamma/\pi=1/2$ and $E_F/\pi E_0=1$, and (d) $\gamma/\pi=0$ and
$E_F/\pi E_0=3/2$. The conductance depends sensitively on the
direction of the magnetization of the two ferromagnets. It is easily
seen that for the fixed $\delta$ doping potential parameters
$\gamma$ and Fermi energy $E_F/\pi E_0$, in Fig.5 (a) and (c) the
conductance is maximum in parallel configuration ( $\theta=0$) while
it is minimum in antiparallel configuration ($\theta=\pi$), which
similar to the conventional magnetoresistance effect. In Fig. 2 (b)
and (d), the conductance takes minimum at the parallel configuration
( $\theta=0$)while it takes maximum near antiparallel configuration
( $\theta=\pi$), which is in stark contrast to the conventional
magnetoresistance effect. From these figure, we can see that the
half-wave loss also exists when the electron wave enter through the
antiparallel configuration according to the analysis above. To
understand these results intuitively, we describe the underlying
physics in Fig. 6 where the spin orientation of incident( red
arrows), transmitted (black arrows) and reflected (blue arrows) are
shown. It is well known [1,4,6] that the electron spin is locked in
the surface plane in  the case of 2D. The spin polarization averaged
in the spin space can be shown , $\langle\sigma_{x(y,z)}\rangle$.
where the normalized wave functions are chosen using the Eq. (2) and
(3). It is easily seen that the spin polarization obviously depend
on the direction of the magnetization. Thus the conductance depends
sensitively on the direction of the magnetization of the two
ferromagnets, which originate from the control of  the spin flow due
to spin-momentum locked.

\begin{figure}
\includegraphics[scale=0.8]{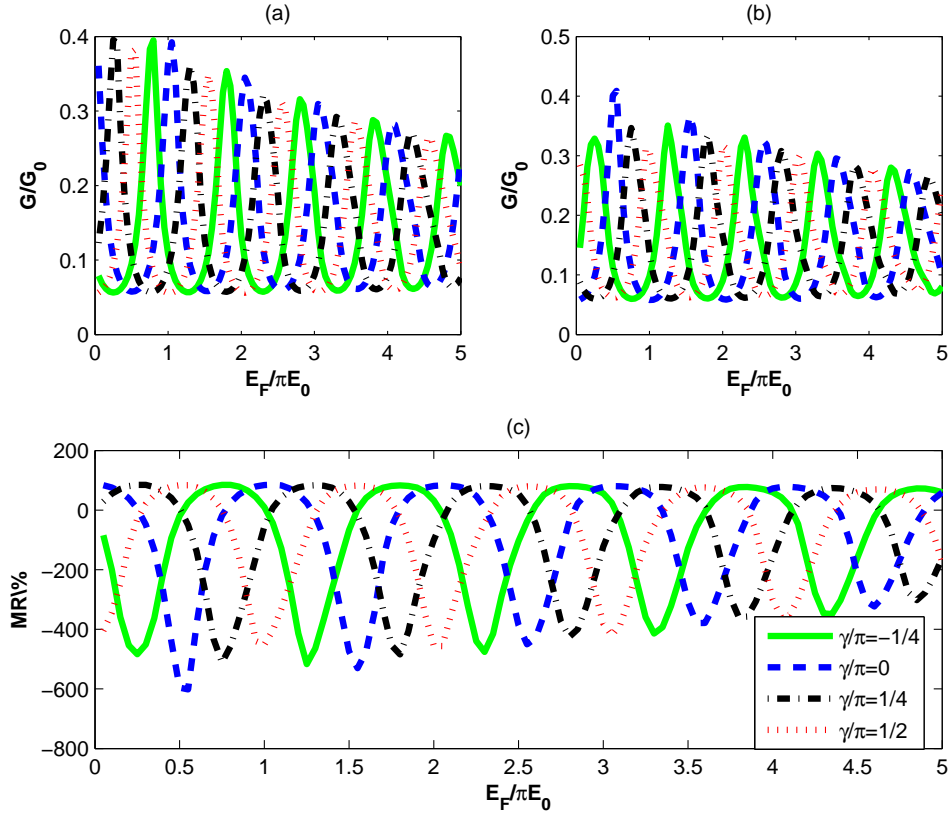}
\caption{Schematic diagram of the normalized conductance [(a) and
(b)] $G/G_0$ and (c) \textbf{MR} as a function of the Fermi energy
$E_F$ with $m_z=m_2=0.9E_F$ and $a=0.5$ for the  different doping
potential parameters  $\gamma/\pi=-1/4, 0, 1/4, 1/2$. The panel (a)
 corresponds to the parallel configuration, and the panel
 (b) corresponds to the antiparallel configuration.}\label{FIG.9.eps}
\end{figure}

\begin{figure}
\includegraphics[scale=0.8]{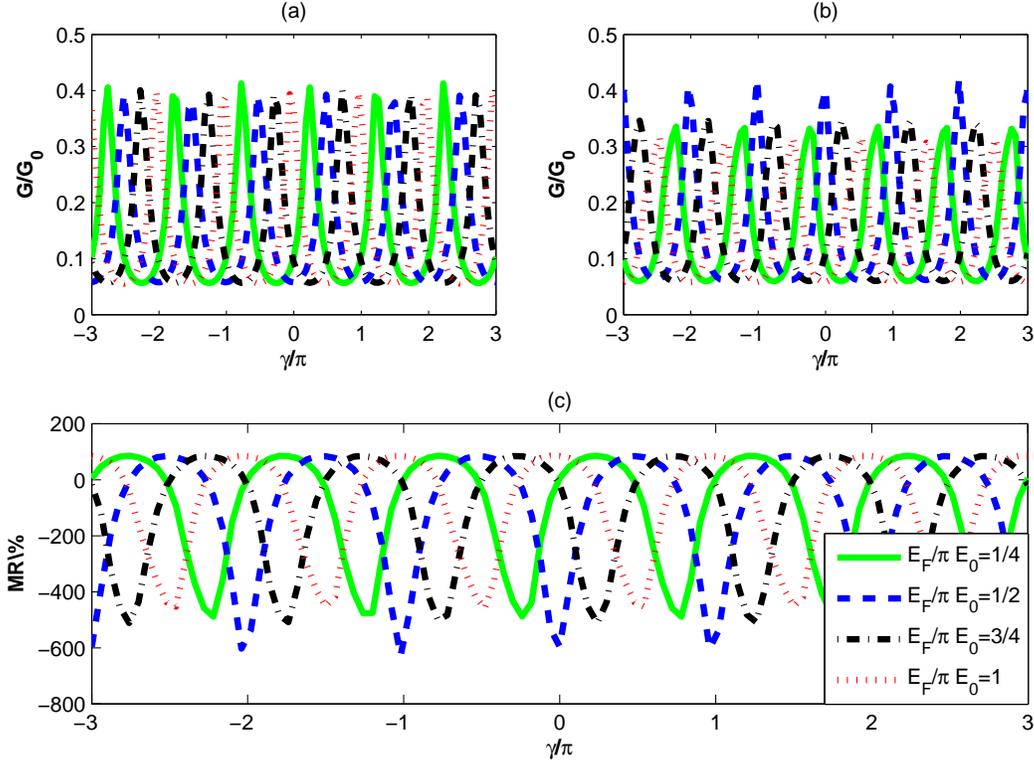}
\caption{Schematic diagram of the normalized conductance [(a) and
(b)] $G/G_0$ and (c) \textbf{MR} as a function of the doping
potential parameters  $\gamma$ with $m_z=m_2=0.9E_F$ and $a=0.5$ for
the  different Fermi energy $E_F/\pi E_0/=1/4, 1/2, 3/4, 1$. The
panel (a)
 corresponds to the parallel configuration, and the panel
 (b) corresponds to the antiparallel configuration.}\label{FIG.10.eps}
\end{figure}

In Fig.7, we show schematic diagram of the normalized conductance
[(a) and (b)] $G/G_0$ and (c) \textbf{MR} as a function of the Fermi
energy $E_F$ with $m_z=m_2=0.9E_F$ and $a=0.5$ for the  different
doping potential parameters  $\gamma/\pi=-1/4, 0, 1/4, 1/2$. The
panel (a)
 corresponds to the parallel configuration, but the panel  (b) corresponds to
  the antiparallel configuration. After obtaining the conductance in the parallel
  configuration ($G_P$) and the conductance in the antiparallel
  configuration ($G_{AP}$),we can define \textbf{MR} as the form:
  $MR=(G_P-G_{AP})/G_{P}$.   Both the conductance and \textbf{MR}
  oscillate  with a $\pi$ period as the Fermi energy increases. It is
easily seen that the \textbf{MR} could be negative as large as 600\%
and the maximum of MR can approach 100\%. That is to say, the
electron conductance obviously change between the parallel
configuration and  the antiparallel  configuration.  It is the
reason that the giant magnetic resistance effect will produce. From
this figure, we can easily see that when the doping potential
parameters $\gamma/\pi>0$, the peak of MR increases and the resonant
peaks of the \textbf{MR} obviously shift to the higher Fermi energy
with increasing of the  doping potential parameters $\gamma/\pi$. We
also can find that by increasing the Fermi energy, the MR
oscillatorily decreases and it has the smaller oscillatory
magnitude. When  the doping potential parameters $\gamma/\pi<0$, the
peak of MR increases , then decreases  with increasing of the Fermi
energy. In Fig.8, the variation of the conductance and the
\textbf{MR} are similar to the Fig.7. It is easily seen that the
oscillatory magnitude is not change in the negative $\gamma/\pi$ and
the positive $\gamma/\pi$, respective. It is because that both the
two region the conductance changes  with a $\pi$ period. It is also
seen that the \textbf{MR} could be negative as big as 600\% and the
maximum of \textbf{MR} can approach 100\%. These characters are very
helpful for making new types of \textbf{MR} devices according to the
practical applications.

\section{Conclusion}
In this paper, We have studied the  electronic transport properties
of the two-dimensional ferromagnetic/normal/ferromagnetic tunnel
junction on the surface of a three-dimensional topological insulator
with taking into $\delta$ doping account. It is found that the
conductance oscillates with the Fermi energy, the position and the
aptitude of the $\delta$ doping. Also the conductance depends
sensitively on the direction of the magnetization of the two
ferromagnets, which originate from the control of  the spin flow due
to spin-momentum locked. It is found that the conductance is the
maximum at the parallel configuration while it is minimum at the
antiparallel configuration and vice versa, which may originate from
the half wave loss due to the electron wave entering through the
antiparallel configuration. Furthermore, we report anomalous
magnetoresistance based on the surface of a three-dimensional
topological insulator.  The \textbf{MR} could be negative as large
as 600\% and the maximum of MR can approach 100\%. That is to say,
the electron conductance obviously change between the parallel
configuration and  the antiparallel  configuration. These characters
are very helpful for making new types of magnetoresistance devices
due to the practical applications.

\vskip0.5cm

\textbf{Acknowledgments}: One of authors(J. H. Y) acknowledges
discussions with Shi Yu (Guangxi medical university, Nanning,
Guangxi, 530021, China ). 

\end{document}